\documentstyle[12pt,epsfig]{article}
\begin{document}
\newcommand{\mat}[4]{\left(\begin{array}{cc}{#1}&{#2}\\{#3}&{#4}
\end{array}\right)}
\newcommand{\nc}{\newcommand}
\nc{\renc}{\renewcommand}
\def\dfrac#1#2{{\displaystyle\frac{#1}{#2}}}
\def\simleq{\; \raise0.3ex\hbox{$<$\kern-0.75em \raise-1.1ex\hbox{$\sim$}}\; }
\def\simgeq{\; \raise0.3ex\hbox{$>$\kern-0.75em \raise-1.1ex\hbox{$\sim$}}\; }
\newcommand{\be}[1]{\begin{equation}\label{#1}}
\newcommand{\beq}{\begin{equation}}
\newcommand{\ee}{\end{equation}}
\newcommand{\beqn}[1]{\begin{eqnarray}\label{#1}}
\newcommand{\eeqn}{\end{eqnarray}}
%
\begin{titlepage}
\begin{flushright}
DFPD/99/TH/45 \\
\today
\end{flushright}

\vspace{2cm}

\begin{center}
{\Large \bf Macroscopic Forces driven by  \\
\vspace{0.3cm}
Resonant Neutrino Conversion}\\

\vskip0.7cm

{\large \bf Dario Grasso\footnote{
E-mail address: grasso@pd.infn.it }
and Anna Rossi\footnote{
E-mail address: arossi@pd.infn.it }
}

\vskip0.5cm
Dipartimento di Fisica, Universit\`a di Padova and
INFN, Sezione di Padova, I-35131 Padova, Italy.
\end{center}
%
\vskip2.cm
\begin{abstract}
\noindent
We show that neutrino oscillations in matter are always accompanied
by collective forces on the medium. This effect may produce
interesting consequences for the background and the neutrino oscillations 
themselves. The force is maximal in the case of resonant neutrino conversion 
in the adiabatic regime. We study here the forces driven by
$\nu_e-\nu_{\mu,\tau}$ and $\nu_e-\nu_s$  MSW conversion and shortly discuss
their possible relevance for the dynamics of a type II supernova.
\end{abstract}
\end{titlepage}
\vskip1.cm
It is well known that the neutrino propagation in matter can be
drastically altered if neutrinos carry non-standard properties
like non-vanishing mass, mixing, magnetic moment, and new interactions
with the matter constituents and the classical fields. The MSW neutrino
resonant conversion \cite{MSW} and the resonant spin-flavour
precession (RSFP)\cite{RSFP} are examples of phenomena that can
take place due to the interplay of the coherent interaction of
neutrinos with the matter and the anomalous neutrino properties.
Usually, the study of such phenomena is focused on the fate of the
neutrinos disregarding any possible implications for the medium
itself. However, such an attitude could be not always justified
especially in those cases where the neutrino flux is very intense,
like during a type II supernova (SN) explosion \cite{Raffelt}.

Recent studies \cite{HarMel,Bing} showed that the coherent
interaction of neutrinos with the matter constituents can gives
rise to macroscopic forces even in the case in which neutrinos are
massless and they have only standard interactions. Such forces are
usually called {\it ponderomotive forces} in analogy to the forces
that arise in an electron gas in the presence of a non-uniform
radiation field \cite{Kibble,LLP}. A similar phenomenon takes
place for the electrons, and other weakly interacting matter
constituents, in the presence of nonuniform flux of
neutrinos. According to the approach of Hardy and Melrose
\cite{HarMel}, the background of neutrinos gives rise to a
space-dependent self-energy, hence to a space-dependent mass
correction, of the matter particles. By interpreting the
statistical average of the mass correction as an interaction
energy density $\mathcal U$, Hardy and Melrose deduced the
existence of a ponderomotive force per unit volume
${\boldmath{\mathcal F}} = -{\boldmath{\nabla}} {\mathcal U}$ which is
proportional to the Fermi constant $G_F$. The reader should not
confuse the ponderomotive force with the more conventional force
produced on the matter constituents by the incoherent elastic
scattering off the neutrinos, which is proportional to $G_F^2$.
In the past years some work has been done concerning collective forces 
produced by coherent neutrino scattering in relation to possible detection 
of relic neutrinos \cite{Langack}. On the basis of the Born
approximation and the assumption of spatially homogeneous neutrino flux,
it was proved that to first order in $G_F$ the effect is vanishing 
\cite{Langack}. We note, however, that these arguments do not apply to
objects of astrophysical sizes where the neutrino flux is not homogeneous.

The relevance of neutrino induced  ponderomotive forces for the
physics of type II supernovae and other extreme astrophysical objects 
was investigated by several authors \cite{HarMel,Bing,Espo,Silva}. 
In some cases, non-negligible effects were predicted. 

The aim of this Letter is to show that neutrino oscillations in
matter are always accompanied by collective forces on the background medium.
Furthermore, we will show that in some phenomenological interesting cases,
these forces are quite larger than the ponderomotive forces 
generated in the absence of neutrino oscillations.   
After deriving the general expression of
the force produced by neutrino oscillations we shall focus
on the case of MSW neutrino conversion. As an
application, we shall briefly discuss the possible relevance of our
results for the dynamics of a type II SN. 

Before entering into the details of our derivation, we have to
mention that in the last few years some debate was going on
concerning the correct expression of the ponderomotive force in
the absence of neutrino oscillations \cite{comments}. Since our
results might be only marginally affected by the outcome of this
controversy, we do not enter here into such discussion. We observe only
that our general treatment lead to results that are consistent
with those of Hardy and Melrose.

Similarly to what done by other authors, we determine the
ponderomotive force per unit volume by taking the gradient of the
interaction energy density $\mathcal U$. The way we compute
$\mathcal U$ is however more straightforward than that followed in
Ref.\cite{HarMel} and it allows us to apply the formalism which is
commonly used to treat neutrino oscillations in matter.
For the sake of
clarity we assume here that the relevant components of the medium
 are well described  by some equilibrium distribution function.
 We also assume that at the position ${\bf x}$, where the force is
computed, neutrinos are already thermally decoupled from the
background. According to the Liouville theorem neutrinos can still be
described by a Fermi-Dirac distribution function
$f_{\nu_a}({\mathbf{x; k}})$ (where the variable ${\bf x}$ enters as a 
parameter) with an effective temperature which will depend on the distance 
from the decoupling position. The
interaction energy density of the neutrinos with the $i-th$ medium
component is computed by taking the thermal average of the
interaction Hamiltonian over the fermion distributions. We have
\begin{equation}\label{inten}
  {\mathcal{U}}_i({\mathbf{x}}) = \sum_{a,b= e,\mu,\tau} \int
\frac {d^3k}{(2\pi)^3} f_{\nu_a}({\mathbf{x; k}}) P_{\nu_a \nu_b}
({\mathbf{x, k}})
   V_{\nu_b i}(\mathbf{x, k})~.
\end{equation}
The quantity
\begin{equation}\label{poten}
  V_{\nu_ai} = \langle \Psi_i \vert {\mathcal{H}}_{\nu_ai} \vert \Psi_i
  \rangle_T
\end{equation}
is the $\nu_a$-potential in a medium composed only by the $i$-th matter
component. In the above ${\mathcal H}_{\nu_a i}$ is the
interaction Hamiltonian density and $\langle \Psi_i \vert ...
\vert \Psi_i \rangle_T$ stands for the thermal average over the
$i$-th particle component of the plasma. It should be noted by the
reader that, in order to account for neutrino oscillations, in the
equation (\ref{inten}) the integral over the neutrino momentum
contains the probability
$P_{\nu_a \nu_b}(\mathbf{x, k}) \equiv P (\nu_a \rightarrow
\nu_b)(\mathbf{x, k}) $.
As we assume isotropic distributions for the neutrinos and
the matter components and we disregard possible effects due to the
polarization of the medium, it is easy to verify that in the
one-loop approximation $V_{\nu_a i}$ does not depend on the
neutrino momentum $\mathbf k$. In this approximation we find that
the force per unit volume produced by all neutrino species on the
$i$-th matter component is
\begin{eqnarray}
\label{force}
  {\bf{\mathcal F}}_i({\mathbf x}) &=& -{\bf \nabla}
  {\mathcal U}_i({\mathbf x})  \nonumber \\
&=& {\mathbf{\mathcal F}}^{osc}_i(\mathbf{x}) - 
  \sum_{a,b= e,\mu,\tau}\!\!
 \!\! \int\!\! \frac {d^3k}{(2\pi)^3}
  P_{\nu_a \nu_b}({\mathbf x,k}) {\mathbf \nabla}
\left[f_{\nu_a}({\mathbf x;k}) V_{\nu_b i}(\mathbf{x})
\right] \nonumber
\end{eqnarray}
where
\begin{equation}\label{osc-force}
 {\mathbf{\mathcal F}}_i^{\mathrm{osc}}({\mathbf x}) =
 -\!\sum_{a,b=e,\mu,\tau}\!\!\! V_{\nu_bi}({\mathbf x}) \!\!
\int \!\!\! \frac {d^3{\mathbf k}}{(2\pi)^3}f_{\nu_a}({\mathbf x; k})
 {\mathbf \nabla}P_{\nu_a\nu_b}({\mathbf{x, k}}).
\end{equation}
The latter contribution
${\mathbf{\mathcal F}}_i^{\mathrm{osc}}$ is the new term induced
by  the oscillations, or  conversions, of a neutrino species into
another -- active or sterile. Clearly, that expression for
${\mathbf{\mathcal F}}_i^{\mathrm{osc}}$ in Eq. (\ref{osc-force})
accounts for whatever kind of neutrino conversion once the proper
probability is specified. We note that in the absence of
neutrino oscillations our expression (\ref{force}) reproduces the
results of Hardy and Melrose \cite{HarMel,comments}.

We predict several interesting effects of the ponderomotive
force induced by neutrino oscillations.
First of all, this force will produce a rearrangement in the density
of the different matter components. Such an effect will be
maximal in the case of MSW (or RSFP) neutrino resonant conversion.
Indeed, in the resonance layer the neutrino survival
probability undergoes a rapid variation which may give rise to
strong macroscopic forces. For this reason, this is the case that we are 
going to discuss in more details here.
As a secondary effect, the density modulation produced by the force will 
affect the neutrino propagation and the neutrino oscillations/conversion.
We shall not study this second order effect here. However, it is worthwhile 
to observe that as a consequence of this effect and of its feedback on the
ponderomotive force, non-linear effects appear which may lead to a
sizeable energy transfer from the neutrino to the plasma \cite{Silva}
and to the amplification (or the fast damping) of the neutrino oscillations.
     
{ \bf 1.}{\it Ponderomotive force due to MSW resonant conversion}.\ \
It is worth recalling the main features of the resonant neutrino
conversion.
We consider the system of two neutrinos  $\nu_e$ and
$\nu_x$ ($x=\mu, \tau, s$) (here $\nu_s$ denote an iso-singlet state),
characterized by
the difference of the eigenstate mass squares $\delta m^2
\equiv m^2_2 - m^2_1$ and the vacuum mixing angle $\theta$.
The coherent neutrino scattering off matter constituents can be described
in terms of matter potentials  as shown in Eq. (\ref{poten}).
In the rest frame of the unpolarized matter, they read as:
\beqn{v_ai}
&&V_{\nu_e e^\mp} =  \pm \frac{G_F}{\sqrt2 m_n} (4 \sin^2\theta_w +1 )\rho
Y_{e^\mp},\\
&&V_{ \nu_{\mu,\tau} e^\mp} =
\pm \frac{G_F}{\sqrt2 m_n} (4 \sin^2\theta_w -1 )\rho
Y_{e^\mp},\\
&&V_{\nu_{a} p} =  \frac{G_F}{\sqrt2 m_n} (1- 4 \sin^2\theta_w )\rho Y_p~, \\
&&  V_{\nu_{a} n} =  \frac{G_F}{\sqrt2 m_n} \rho Y_n, ~~~~~~
V_{\nu_{s} i} = 0 ~, ~~~~~~~~~ i = e^\mp, p, n
\eeqn
here $\rho$ is the matter density, $Y_i$ the concentration of the
$i$-type component and $m_n$ the nucleon mass (for the
anti-neutrinos, the above potentials change sign). The Hamiltonian
${\bf H}$  governing the evolution equations in matter can be
written as:
\beqn{Hm}
 {\bf H} &=&
 \mat{\frac{\delta m^2}{2 E} \cos 2\theta -
\Delta V} {\frac{\delta m^2}{4 E} \sin 2\theta} {\frac{\delta
m^2}{4 E} \sin 2\theta}{0}~,
\nonumber \\
 \Delta V &=& \sum_i
(V_{\nu_ei} -V_{\nu_x i}) = \sqrt2\frac{G_F}{m_n}\rho_{eff}(r)
\eeqn
here $\Delta V$ is the effective matter potential for the
system of $\nu_e -\nu_x$ and $\rho_{\mathrm{eff}}= \rho Y$ where
the net effective concentration $Y$ is $Y= Y_e \equiv Y_{e^-} -
Y_{e^+}$ for the $\nu_e-\nu_{\mu, \tau}$ system and $Y= Y_e
-\frac12 Y_n$ for the $\nu_e -\nu_{s}$ channel.
The  mixing angle $\theta_m$
and the neutrino wavelength $\lambda_m$ in matter are
 given by:
\beqn{matter-p}
&&\sin^2  2\theta_m =\frac{(\delta m^2 \sin 2\theta)^2}{
 (\delta m^2 \cos2\theta - 2 E \Delta V )^2 + (\delta m^2 \sin 2\theta)^2 }~,\\
&& \lambda_m = \frac{\delta m^2\lambda}{
\sqrt{(\delta m^2 \cos2\theta - 2 E \Delta V )^2 +
(\delta m^2 \sin 2\theta)^2 }}
\eeqn
where $\lambda = \displaystyle \frac{4\pi E }{\delta m^2}$ is the vacuum
wavelength.

As it is well known the efficiency  of the
conversion is determined by the resonance condition,
$\dfrac{\delta m^2 \cos 2\theta}{2 E} = \Delta V $
and by  the adiabaticity property $\dfrac{\mathrm {d}\theta_m}{
\mathrm{ dr}}\ll \pi/\lambda_m$ where \beqn{Dtheta}
&&\frac{\mathrm {d}\theta_m}{ \mathrm{ dr}} = \sin^2 2\theta_m ~
\frac{  E \Delta V}{\delta m^2\sin 2\theta } h^{-1}(r) ~, \\
&&h^{-1}(r) \equiv \frac{\mbox{d ln}
(\rho_{\mathrm{eff}})}{\mbox{dr}}~.
 \eeqn
The $\nu_{e}$ survival probability at a certain distance $r$ from the source
is given by
\be{Pnue}
P_{\nu_{e}}(r;E) = \frac12 + (\frac12 -P_{c}) \cos2\theta_m(r_i)
\cos2\theta_m(r)~,
\ee
where $r_i$ is the (radial) coordinate of the neutrino source and the
function $P_{c}$ is
the probability of jumping
from one matter eigenstate to the other \cite{Parke}.
In typical cases, $E\Delta V \gg \delta m^2$ in the neutrino
production region which implies $\cos 2\theta_m(r_i) \approx -1$.
We assume the resonant conversion to be  adiabatic
 --$P_{c} \simeq 0$  (this will be justified later on).
Then, from Eqs. (\ref{Dtheta}) and (\ref{Pnue}) we can easily obtain
\be{gradP}
\frac{\partial P_{\nu_{e}}}{\partial r} = \sin 2\theta_m
\frac{\mathrm {d}\theta_m}{ \mathrm{ dr}} =
\sin^3 2\theta_m \frac{  E \Delta V}{\delta m^2 \sin 2\theta} h^{-1}(r)~.
\ee
We note that the r.h.s. of Eq.(\ref{gradP}) attains a maximum when the
resonance condition is fulfilled ($\sin^2 2\theta_m= 1$).

Now we are ready to turn back to the  expression for the
ponderomotive force induced by the neutrino resonant conversion
given in Eq. (\ref{osc-force}). For simplicity, we  assume that
only $\nu_e$'s are created by some point-like source with a
 thermal distribution $f_{\nu_e}(r;E)$ and  propagate
isotropically trough an inhomogeneous medium like that e.g.
present in a star. Expressing the neutrino density in term of the
luminosity $L_\nu$  ($n_\nu = \dfrac{L_\nu}{4 \pi r^2 \langle
E\rangle}$) we find from Eq. (\ref{osc-force})
\begin{eqnarray}
\mathbf{\mathcal{F}}^{osc}_i(r) &=& - (V_{\nu_e i} - V_{\nu_x i})
\frac{L_\nu}{4\pi r^2} \dfrac{\int dE  E^2 f_{\nu_e}(E)
\dfrac{\partial P_{\nu_{e}}(r, E)}{\partial r} } {\int dE  E^3
f_{\nu_e}(E)}\nonumber \\ & = & - (V_{\nu_e i} - V_{\nu_x i})
\frac{\Delta V}{\delta m^2 \sin 2\theta} \frac{L_\nu}{4\pi r^2}
h^{-1}(r) I(r)
\end{eqnarray}
where \be{integral}
 {\mathrm{I}}(r) \equiv  \frac{\int dE  E^3
\sin^3 2\theta_m(r,E) f_{\nu_e}(E) } {\int dE  E^3 f_{\nu_e}(E)}~.
\ee
Since the function $\mathrm I$ is obtained by a convolution of
 $\sin^3 2\theta_m(E)$
with the neutrino distribution function, it
attains the maximum value $\mathrm I_*$ at the position  $r_*
\equiv r_{\mathrm{res}}\vert_{E = \langle E\rangle}$,
where $\langle E\rangle$ is the
neutrino mean energy. The actual value of $\mathrm I_*$ depends on
the width $\Delta E \propto \sin 2\theta \dfrac{\delta m^2}{\Delta
V}$ of the function $\sin^3 2\theta_m(E)$, hence on the degree of
adiabaticity of the neutrino conversion. Therefore the maximal
force is \be{Fres}
 {\mathbf{\mathcal F}}_i^{\mathrm{osc}}(r_*) =
 - (V_{\nu_e i} - V_{\nu_x i})\frac{L_\nu}{8\pi r^2 \langle E\rangle}
 h^{-1}(r_*)\frac{I_*(\theta)}{\tan 2\theta} ~.
\ee
By comparing  this expression with that of the conventional force
driven by the gradient of the neutrino density (i.e. the second
term in eq. ({\ref{force}), hereafter named ${\mathcal F}^{HM}$)
we note a significative difference. 
While the latter is
proportional to the net neutrino density $N_\nu$, the former
depends only on the helicity species which undergoes the resonant
conversion. Such a different behavior may play a crucial role in
SNs and in other astrophysical or cosmological environments.
Upon  making the comparison more quantitatively,
we  consider the forces produced on the
electron  component of the medium.
\begin{figure}[t]
\vskip -1cm
\centerline{\protect\hbox{
\psfig{file=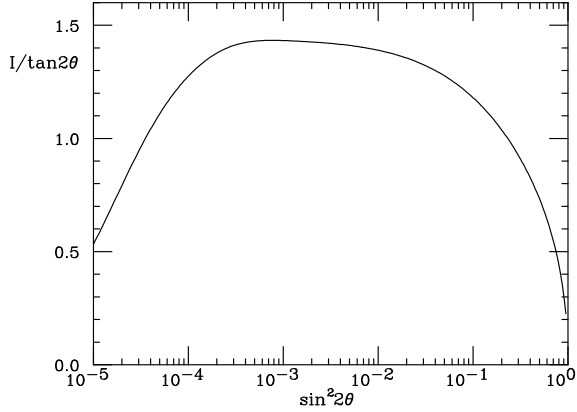,
width=9.3cm,angle=90}}}
\vskip -1.0cm
\caption{The
ratio $I_*(\theta)/{\tan 2\theta}$ is plotted as a function of
$\sin^22\theta$.}
\end{figure}
We find
\be{MSWvsHM}
\left. \frac{{\mathcal F}_e^{\mathrm{osc}}} {{\mathcal F}^{HM}}
\right|_{r_*} \approx
\frac{r_*h^{-1}_*}{4\xi} \frac {I_*(\theta)}{\tan 2\theta} ~,
\ee
where the parameter $\xi \equiv 1 - \frac {\langle E\rangle }
{\langle \bar E\rangle}$
accounts for the possible difference in the $\nu_e$ and
$\bar\nu_e$ spectra (we assume that $L_{\nu_e} = L_{\bar\nu_e}$).
In the Fig.1 we plot the function $I_*(\theta)/{\tan 2\theta}$.
We observe that this function does not lead to any suppression in
the phenomenological interesting range $ 10^{-4} \simleq
\sin^2 2\theta \simleq 2 \times 10^{-1}$.
The actual value of ${r_*}{h}^{-1}$ depends on the matter density
profile and on the kind of neutrino transition (active-active or
active-sterile) considered. Below we will show as in a SN such a
quantity typically ranges from a values of few units to some
powers of $10$. We should keep in mind, however, that the
adiabaticity requirement implies an upper limit to the allowed
value of $h^{-1}$. This is given by
\beqn{adiab}
 h^{-1} \ll  h_{\mathrm{adiab}}^{-1} =
\frac{\delta m^2 \sin^2 2\theta}{2 \langle E \rangle \cos \theta}~.
\eeqn
 Above this value the scale setting the gradient of $P_{\nu_e}$ is
$\lambda_m$  which, in the non-adiabatic regime, is larger than
$h$. Therefore the maximal force is achieved in the adiabatic
regime.

For completeness, we also compare the force
$\mathbf{\mathcal{F}}_e^{\mathrm{osc}}$ with the outward force due
to the elastic scattering off electrons, ${\mathbf{\mathcal
F}}_e^{\mathrm{scatt}}\sim \dfrac{G^2_F}{2\pi^2}
\dfrac{L_\nu}{4\pi r^2} \langle E\rangle T_e n_e$. We find
 \be{MSWvsScatt}
\left. \frac{\mathbf{\mathcal{F}}_e^{\mathrm{osc}}}
{F^{\mathrm{scatt}}_e}\right|_{r_*}\approx 8\times 10^{-7} 
\left(\frac{1\mathrm{MeV}}{T_e}
\frac{100\mathrm{MeV}^2}{E^2}\frac{h^{-1}}{10^{-5}\mathrm{cm}^{-1}}
\right) \frac {I_*(\theta)}{\tan 2\theta}
\ee
 where $T_e$ is the electron temperature. Clearly
$\mathbf{\mathcal{F}}_e^{\mathrm{osc}} \ll {\mathbf{\mathcal
F}}_e^{\mathrm{scatt}}$. We observe, however that these forces are
qualitatively quite different and cannot be always directly
compared. For example, the ponderomotive force naturally excite
acoustic and plasma waves which can hardly be done by the force due to
the incoherent neutrino scattering. Therefore the former may induce effects 
which are not produced by the latter allowing its possible 
identification.

{\bf 2.}{\it  Ponderomotive force in supernova}.\ \  As an
application of our previous results, we now investigate what kind
of effects the ponderomotive forces produced by neutrino resonant
conversion may give rise to during a type II SN explosion. We
consider the region above the neutrino sphere at a time after the
core bounce. In this epoch all the neutrino (as well
anti-neutrinos) species are emitted with approximately the same
luminosity. However the individual neutrino energy distributions
may be quite different. For this reason the force produced, say by
the $\nu_e \rightarrow \nu_\mu$ MSW conversion is not canceled by
the opposite force produced by $\nu_\mu \rightarrow \nu_e$ that
will occur more externally. In both cases the resonance takes
place in a dynamically relevant region for $\delta m^2 = 10 \div
10^4~\mathrm{eV}^2$ if $\langle E\rangle \simeq 10$ MeV. In the
earlier epoch ($t \simleq 1$ s) with a typical profile
\cite{Bethe} $N_e = 10^{34} r_7^{-3}~ \mathrm{cm}^{-3}$ ($r_7
\equiv r/10^7~\mathrm{cm}$), from eq.(\ref{MSWvsHM}) we find
$\left.\dfrac{\mathbf{\mathcal{F}}_e^{\mathrm{osc}}} {\cal
F^{HM}}\right|_{r_*} \sim 1$ with $\xi \sim 0.3$. At later times, as
the scale $h$ becomes smaller, this ratio can increase up to an
order of magnitude.

A different picture may emerge for the $\nu_e-\nu_s$, or the
${\bar \nu}_e-{\bar \nu}_s$, channels. We recall that in this case
$\rho_{\mathrm{eff}} = \rho (3Y_e - 1)$ and the resonance
condition can be satisfied for arbitrarily small values of $\delta
m^2$.
Hence, for small values of $\delta m^2$ the resonance take place
nearby the point where the potential approaches zero. As a
consequence $h^{-1}(r_*)$ could be quite large. At the same time,
however, the adiabaticity requirement (\ref{adiab}) set an upper
limit to  $h^{-1}$  which is proportional to $\delta m^2$. Adopting
standard SN density profile \cite{NPRV} and assuming $\sin^2
2\theta \simleq 10^{-2}~$, we found that the maximal force is
obtained for $\delta m^2 \sim 1~\mathrm{eV}^2$,  corresponding to
$\Delta V \simeq 10^{-7}$ eV and $h^{-1} \sim
10^{-5}~\mathrm{cm}^{-1}$ ($r_* \simeq 150$ km). Therefore the
ratio (\ref{MSWvsHM}) can be as large as $10^2$.

A comment at this point is in order. Because the electron neutrino
potential changes sign at the radius where $Y_e = 1/3$, two kinds
of resonance should take place close to this position. If we
assume $ 0 < \delta m^2\simleq 10^2 ~\mathrm{eV}^2$,
and since $Y_e$ is a growing function of
the radius above the neutrino-sphere, the transition ${\bar
\nu}_e-{\bar \nu}_s$ will first occur in the region where  $Y_e <
1/3$. It is easy to verify that in this case the force pulls
electrons inwards. Soon afterwards the transition ${\nu}_e-{
\nu}_s$ take place in the region with $Y_e > 1/3$ and thereby the
force goes in the opposite direction. It happens \cite{NPRV} that
the two types of conversion take place very
close to each other. However the corresponding forces should not
cancel as the average energy of $\nu_e$ and ${\bar \nu}_e$ differs
of some 30-50\%.  In the case $\delta m^2 > 10^2 ~
{\mathrm eV}^2$ only the conversion ${\bar \nu}_e-{\bar \nu}_s$
can happen as far as $Y_e< 1/3$ \cite{NPRV}.

It is interesting to estimate the velocity which background matter
close to the resonance position acquires under the action of
$\mathbf{\mathcal{F}}_e^{\mathrm{osc}}$ in a time interval of the
order of the dynamical time $\simeq 0.1$ s. A straightforward
computations shows that velocities as large as $\sim 10^2~\mathrm{km/s}$ 
can be reached. Other interesting effects arise if these
ponderomotive forces induce plasma instabilities \cite{Silva} which may 
give rise to a significative energy transfer from the neutrino to
the plasma.

In conclusion, we think that the phenomenological perspectives
open by the study of the ponderomotive force produced by neutrino
oscillations are rich and deserve further investigation.

\vspace{1.cm}

\noindent
{\bf Acknowledgements} We thank M.~ Pietroni for helpful discussions.

\end{document}